\documentstyle[prb,aps,twocolumn,epsf]{revtex}
\begin{document}
\draft
\title{ 
Electronic structures of antiperovskite superconductors: MgXNi$_3$ (X=B,C,N)
}
\author{ J. H. Shim and B. I. Min}
\address{ Department of Physics, Pohang University of Science and
Technology, Pohang 790-784, Korea}
\maketitle
\begin{abstract}
We have investigated electronic structures of 
a newly discovered antiperovskite superconductor MgCNi$_3$ and 
related compounds MgBNi$_3$ and MgNNi$_3$.
In MgCNi$_3$, a peak of very narrow and high density of states 
is located just below $\rm E_F$, which corresponds to the $\pi^*$ 
antibonding state 
of Ni-$3d$ and C-$2p$ but with the predominant Ni-$3d$ character. 
The prominent nesting feature is observed in the $\Gamma$-centered
electron Fermi surface of an octahedron-cage-like shape
that originates from the 19th band. 
The estimated superconducting parameters based on the simple rigid-ion 
approximation are in reasonable agreement with experiment, 
suggesting that the superconductivity in MgCNi$_3$ is described well by
the conventional phonon mechanism.
\end{abstract}

Recently, He {\it et al.}\cite{he} have discovered a new intermetallic 
superconductor MgCNi$_3$ with the transition temperature T$_C$ of 8K, 
which has the antiperovskite structure.
Because it has a large proportion of Ni per unit cell, it is expected that 
the magnetic fluctuation would be important in determining the 
superconducting behavior.
This system is reminiscent of another Ni-based superconductors
LnNi$_2$B$_2$C (Ln = Y, Tm, Er, Ho, Lu) \cite{Naga,cava,pickett}.
The band calculations indicate that very large and narrow energy peak 
in the density of states (DOS) is located just below the Fermi energy $\rm E_F$
which has mainly the Ni 3d character\cite{hayward,Dugdale}.
The behavior of the upper critical field $H_{c2}(T)$ can be well
fitted with the conventional BCS expression \cite{li},
whereas the zero-bias conductance peak observed below T$_C$ suggests 
that MgCNi$_3$ is likely to be a non-s wave superconductor\cite{mao}.

With Cu doping (electron doping) on Ni-site, T$_C$ decreases systematically, 
but with Co doping (hole doping), the superconductivity disappears 
abruptly for doping of only 1\% \cite{hayward}.
In the case of Co doping, there is no evidence that the quenching 
of superconductivity is related to magnetism.
Furthermore, it is observed that the superconductivity of MgC$_x$Ni$_3$ 
is sensitive to the content of C; it disappears 
between x=0.96 and x=0.90 \cite{he}.

To understand the mechanism of the superconductivity in MgCNi$_3$, 
we have investigated systematically the electronic structures of 
MgXNi$_3$ (X=B, C and N),
which have different number of valence electrons, but have the similar 
band structures. Using the linearized muffin-tin orbital (LMTO) band
method in the local density approximation (LDA), we have obtained band 
structures, DOSs, and Fermi surfaces, and discussed the bonding characters. 
Muffin-tin orbitals up to $d$-states for Mg, C, and
up to $f$-states for Ni are included in the LMTO band calculations.
We have also estimated superconducting parameters based on the rigid-ion 
approximation.
We have considered cubic MgCNi$_3$ with the lattice constant of 3.81\AA
\cite{he} and employed the atomic radii of  3.20, 1.54 and 2.49 \AA~ 
for Mg, C and Ni, respectively. The same structural parameters are
used for all MgXNi$_3$.

MgCNi$_3$ has the cubic antiperovskite structure: Mg at (0 0 0), 
C (0.5 0.5 0.5), and Ni at (0.5 0.5 0), (0.5 0 0.5) and (0 0.5 0.5). 
It is called as an antiperovskite
structure because the transition metals are located at the corners of the 
octahedron cage in contrast to the ordinary perovskite structure \cite{Papa}. 
Without C located at the center of cubic cell, MgNi$_3$ is a simple ordered 
intermetallic compound with fcc structure ($Cu_3Au$-type). 
Without C, the Ni-$3d$ band of MgNi$_3$ is very narrow, leading to
a magnetic ground state with Ni magnetic moment of 0.43$\mu_B$ \cite{shim}.
By inserting C, two carbons become nearest neighbors of Ni, and
thus it is expected that Ni-$3d$ and C-$2p$ electrons are strongly hybridized.

\begin{figure}
\epsfxsize=200pt
\epsffile{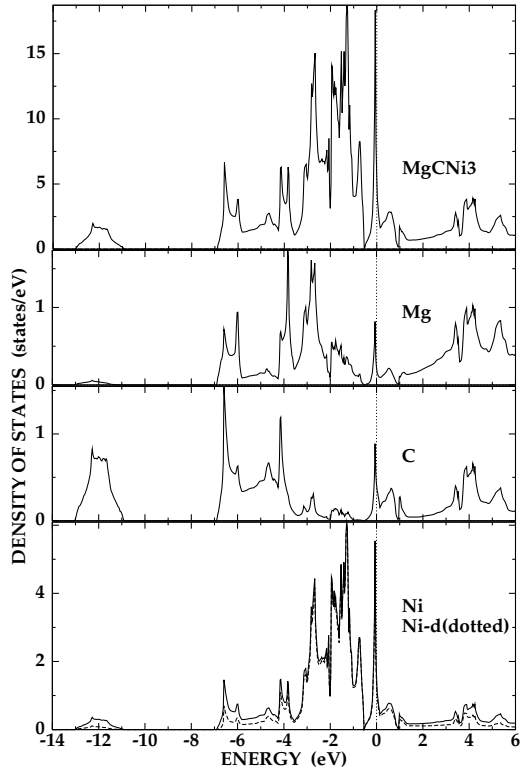}
\caption{ Total and projected local DOS of MgCNi$_3$.}
\label{dos}
\end{figure}

Total and projected local DOS of MgCNi$_3$ are provided in Fig.\ \ref{dos}. 
The overall shape of the total DOS is similar to those of Hayward {\it et al.}
\cite{hayward} and Dugdale and Jarlborg \cite{Dugdale}.
The peaks near $-$7eV and 4eV correspond to $\sigma$ bonding and
antibonding states, respectively, of Ni-$3d$ and C-$2p$ states. 
On the other hand, the peaks near $-$4eV 
correspond to $\pi$ bonding states of Ni-$3d$ and C-$2p$.
The $\pi^*$ antibonding states are located just below $\rm E_F$, 
yielding the high DOS at $\rm E_F$ ($N(E_F)$) of 5.34 [states/eV].
The contribution of Ni-$3d$ states to the DOS at $\rm E_F$ is as much as 76\%. 
Small amount of C-$2p$ states are hybridized with Ni-$3d$ states. 
Because the peak just below $\rm E_F$ is very high and narrow, 
this system is expected to be unstable by small perturbation.
The peak is located $\sim 60$meV below $E_F$, and 
about 0.5 electrons are occupied between the peak and $E_F$.
The stoner parameter S, defined as $S\equiv N(E_F)I_{XC}$
with $I_{XC}$ denoting the intra-atomic
exchange-correlation integral, is 0.64. Indeed the filling of holes in 
a rigid band scheme produces the magnetic instability. 
That is, replacing Ni by a virtual atom with atomic number 27.93 
(corresponding to Co 7\% doping) yields the stoner parameter larger 
than 1.00 \cite{shim}.

\begin{figure}
\epsfxsize=200pt
\epsffile{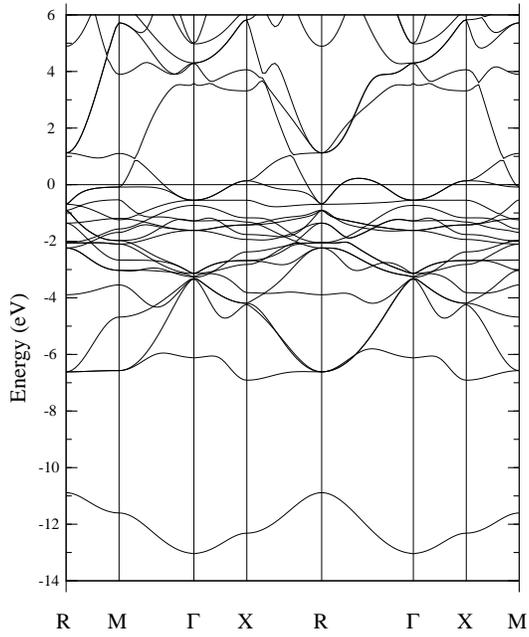}
\caption{ Band structure of MgCNi$_3$ along the symmetry lines of
the simple cubic Brillouin zone}
\label{band}
\end{figure}

The band structure of MgCNi$_3$ along the symmetry line of the simple cubic 
Brillouin zone is shown in Fig.\ \ref{band}. 
The band near $-$12eV corresponds to C-$2s$ states, while
the dispersive bands in the range of $-$7eV to $-$4eV are due
to C-$2p$ states. Only two
bands (the 18th and 19th bands), which have mainly Ni-$3d$ character, cut the
Fermi level. These two bands are confined between $-$0.5eV and 1.0eV,
and available states in these bands are about four. 
Hence, in the rigid band scheme, small electron or hole doping will 
produce the carriers with the 18th and the 19th band character.
The 18th band is relatively flatter than the 19th band, and so the Ni-$3d$ 
character is stronger in the 18th band.

\begin{figure}
\epsfxsize=200pt
\epsffile{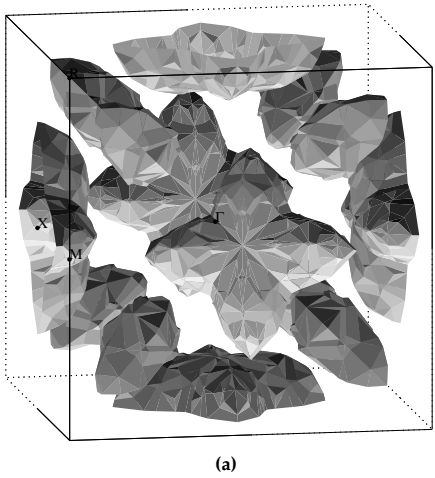}
\epsfxsize=200pt
\epsffile{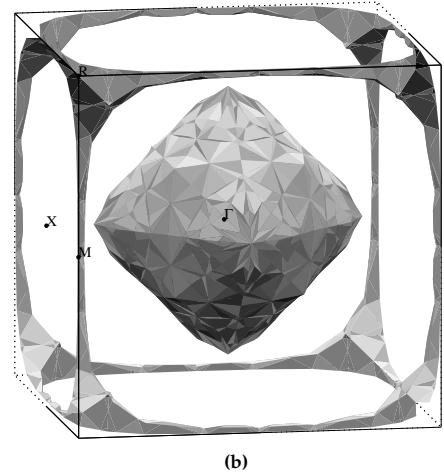}
\caption{ Fermi surfaces of MgCNi$_3$ in the simple cubic Brillouin zone
coming from the 18th band (a) and the 19th band (b)}
\label{fers}
\end{figure}

Figure\ \ref{fers} presents the Fermi surfaces of the 18th (a) and the 
19th bands (b) in the simple cubic Brillouin zone. 
The 18th band gives rise to a clover-like hole surface centered at 
X of each cubic face and small hole pockets along the (111) directions. 
With increasing the Fermi level (electron doping), 
the areas of these hole Fermi surfaces decrease, 
reducing $N(E_F)$.
In opposite, with decreasing the Fermi level (hole doping),
the Fermi surface areas increase, enhancing $N(E_F)$.
When decreasing the Fermi level further below the DOS peak,
the crossing of each hole surface occurs, converting it
to electron surface and decreasing Fermi surface.  
On the other hand, the 19th band yields an octahedron-cage-like 
electron surface
centered at $\Gamma$ and additional narrow electron surfaces 
along the Brillouin zone edges (Fig. \ref{fers}(b)). As compared to the case of the
18th  band, the 19th band shows a rather small change in the Fermi surface 
topology
with varying the Fermi level position due to its more dispersive band character. 
The Fermi surface topology obtained in the present study is qualitatively
similar to those of Dugdale and Jarlborg \cite{Dugdale}.

The most notable in Fig. \ref{fers}(b) is the prominent nesting feature 
along the (110) direction observed in the $ab$ plane of the
$\Gamma$-centered Fermi surface with the octahedron-cage-like shape.
In fact, this is contrary to the report by Dugdale and Jarlborg \cite{Dugdale} who
have not observed the obvious nesting feature in the Fermi surface of the 19th band. 
As mentioned above, the detailed shapes of the Fermi surfaces are very 
sensitive to the position of the Fermi level, because of the very sharp 
DOS peak near $\rm E_F$. Presumably the difference between 
two results arises from the different band parameters employed in the
LMTO band calculations, such as atomic radii, number of k-points,
energy parameters, and so on. In any case, the present result reveals that
the system is in the vicinity of the Fermi surface nesting, if not complete
in undoped MgCNi$_3$. 
It is well known that the system with the Fermi surface nesting 
can be strongly correlated with various instabilities: 
the structural transition (the charge 
density wave instability) or the spin density wave instability. 
It is thus expected that MgCNi$_3$ may be susceptible to one of
the above instabilities. However, until now, any evidence of magnetic or
structure transition has not been reported \cite{hayward,huang}.
This aspect remains to be resolved. 
 
\begin{figure}
\epsfxsize=200pt
\epsffile{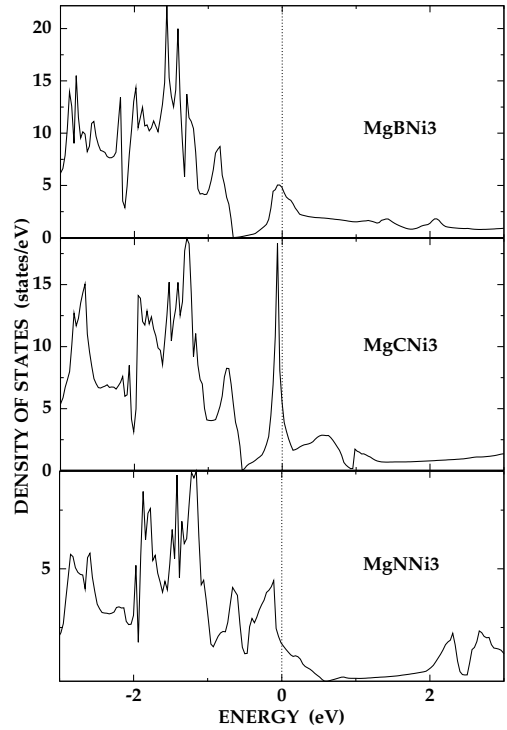}
\caption{ Density of states of MgXNi$_3$ (X = B, C, N) }
\label{dos_comp}
\end{figure}

We have seen that the DOS peak near $\rm E_F$ is produced 
by the hybridization of Ni-$3d$ and
C-$2p$ states. In MgCNi$_3$, the bands from $-$0.5eV to 1.0eV are
almost half-filled: two electron states out of four available 
states are occupied. 
To explore the doping effect with varying the number of valence electrons, 
we have investigated electronic structures of
MgXNi$_3$ (X= B, C, N) . Figure \ref{dos_comp} provides
the DOSs for MgXNi$_3$. It is seen that the effect of changing X is mainly
a variance of the Fermi level with respect to the DOS peak. 
The shape of DOS is perturbed a little, which indicates that the rigid band
scheme would work well in this system.

The B-$2p$ state in MgBNi$_3$ is located higher in energy
than the C-$2p$ state of
MgCNi$_3$, and so the hybridization with Ni-$3d$ is stronger.
Hence the band is more dispersive and accordingly 
the DOS peak becomes smeared.
Although the Fermi level in MgBNi$_3$ is located very close to the DOS peak, 
the DOS at $\rm E_F$, 4.79 [states/eV], is comparable to that of MgCNi$_3$ 
(5.34 [states/eV]) (Table\ \ref{Nl}).  
Hence the magnetic instability does not occur either in MgBNi$_3$. 
As described before, the effective hole doping in MgBNi$_3$ 
converts a hole Fermi surface of the 18th band to an
electron Fermi surface, and an electron surface of the 19th band is
reduced smaller.  
In MgNNi$_3$, the N-$2p$ state is located a bit lower in energy 
than the C-$2p$ state of MgCNi$_3$, yielding reduced bandwidths 
of both the 18th and the 
19th band.  By the effective electron doping in MgNNi$_3$, 
the DOS at $\rm E_F$ is reduced to 3.63 [states/eV] (Table\ \ref{Nl}). 
The contribution of the 18th band
to the DOS at $\rm E_F$ is almost negligible and the Fermi surface of the
19th band, which gives the main contribution to the DOS at $\rm E_F$, 
is changed to a hole surface.
 
We have explored superconducting properties of MgXNi$_3$ based on 
the rigid-ion approximation \cite{gaspari}.
We have estimated the superconducting parameter 
$\eta_{\alpha}=N(E_F)\langle I^2_{\alpha}\rangle$, where
$\langle I^2_{\alpha}\rangle$ is the
average electron-ion interaction matrix element for the 
$\alpha$-th ion. Table \ \ref{eta} provides the 
calculated $\eta_\alpha$ for each MgXNi$_3$. 
It is seen that the contribution of Ni-$3d$ states to
the superconductivity is most important and $\eta_{Ni}$ is
the largest for MgCNi$_3$. This is consistent with the observed
trend that both the electron and hole dopings on MgCNi$_3$
suppress the superconductivity. 
By increasing the atomic number from B, C to N, the contribution of 
X-$2p$ states increases, while that of Ni-$3d$ increases first and then
decreases. 
Due to light ionic masses of X, even the small increase in $\eta_{X}$ 
affects the superconducting property substantially.
Therefore, effectively electron doped system MgNNi$_3$, 
once synthesized successfully in the antiperovskite structure,
would have comparable $T_C$ to MgCNi$_3$. 

One can evaluate the electron-phonon coupling constant
$\lambda_{ph}$ by using the McMillan's formula 
$\lambda_{ph} = \sum_{\alpha}\eta_{\alpha} /M_{\alpha}\langle
\omega_{\alpha}^{2}\rangle$,
where $M_{\alpha}$ is an ionic mass and $\langle\omega_{\alpha}^{2}\rangle$ is 
the relevant phonon frequency \cite{mcmillan}. Since there has been no
information on the relevant phonons, we instead use the 
average phonon frequency $\langle\omega^{2}\rangle\simeq\Theta_{D}^{2}/2$,
where $\Theta_{D}$ is the Debye temperature. However, 
even the value of $\Theta_{D}$ is not available. 
Albeit very crude, one can estimate $\Theta_{D}$
from the specific heat data \cite{he}: $\sim$ 300 K from the graph of C/T vs. 
T$^2$.  Using these informations for MgCNi$_3$, one
obtains $\lambda_{ph}=1.56$, and then the McMillan's T$_C$ formula 
with an effective electron-electron interaction parameter
$\mu^*=0.13$ gives rise to $T_C=23$ K.
These values seem to be too large, as compared to experimental $T_C$ 
and the estimated $\lambda_{ph} \sim 0.8$ 
from the specific heat data \cite{he}. Note, however, that 
$\lambda_{ph}$ strongly depends on the choice of the Debye temperature. 
As shown in Table\ \ref{eta}, with a choice of larger 
$\Theta_{D}=400$ K, one obtains $\lambda_{ph}=0.77$ and $T_C= 11$K, 
which are in reasonable agreement with experiment.
This suggests that the superconductivity in this system 
can be described by the conventional phonon mechanism.
For more precise estimations of $\lambda_{ph}$ and $T_C$, 
detailed information on the phonon spectra is prerequisite.

In conclusion, we have investigated electronic structures of the non-oxide
antiperovskite superconductor MgCNi$_3$. The $\pi^*$ antibonding state of 
Ni-$3d$ and C-$2p$ is formed near $\rm E_F$ with the mainly 
Ni-$3d$ character. Fermi surfaces are composed of two bands. 
The topology of hole Fermi surfaces coming from the 18th band is 
sensitively modified by the variance of the Fermi level position. 
The electron surface of the 19th band with an octahedron-cage shape 
tends to induce the Fermi surface nesting.
By comparison of DOSs for MgXNi$_3$ (X=B,C,N), 
the doping effects are discussed.
The estimation of $\lambda_{ph}$ and $T_c$ based on the rigid-ion
approximation suggests that the superconductivity of MgCNi$_3$ is described 
well by the conventional phonon mechanism.

Acknowledgments $-$ 
This work was supported by the KOSEF through the eSSC at POSTECH
and in part by the BK21 Project. Helpful discussions with N.H. Hur are
greatly appreciated.

%---------------------
%---------------------
%---------------------
%---------------------
%---------------------
\begin{table}
\setdec 0.00
\caption{Total and partial DOSs at $\rm E_F$ (in states/eV)
for MgXNi$_3$ (X = B,C,N).}
\begin{tabular}{lccccc}
 &$ N_{\rm Mg}$  &$ N_{\rm X}$ &$ N_{\rm Ni}$ &$ N_{\rm total}$\\
\tableline
$\rm MgBNi_3$&\dec 0.38 &\dec 0.18 &\dec 1.41 &\dec 4.79 \\
$\rm MgCNi_3$&\dec 0.22 &\dec 0.42 &\dec 1.57 &\dec 5.34 \\
$\rm MgNNi_3$&\dec 0.11 &\dec 0.34 &\dec 1.06 &\dec 3.63 \\
\end{tabular}
\label{Nl}
\end{table}
\begin{table}
\setdec 0.00
\caption{ Comparison of $\eta$ (in eV/\AA$^2$) and $\lambda_{ph}$
 for $\Theta_{D}$=300K and 400K.}
% for MgXNi$_3$(X = B, C and N) }
\begin{tabular}{lccccc}

&$ \eta_{\rm Mg}$  &$ \eta_{\rm X}$ &$ \eta_{\rm Ni}$ &
        $\lambda_{ph}$ (300K) & $\lambda_{ph}$ (400K)\\
\tableline
$\rm MgBNi_3$&\dec 0.00 &\dec 0.22 &\dec 0.67 &\dec 0.67 &\dec 0.38 \\
$\rm MgCNi_3$&\dec 0.00 &\dec 0.48 &\dec 1.36 &\dec 1.36 &\dec 0.77 \\
$\rm MgNNi_3$&\dec 0.00 &\dec 0.58 &\dec 0.87 &\dec 1.07 &\dec 0.60 \\
\end{tabular}
\label{eta}
\end{table}
\end{document}